\documentclass [prb,superscriptaddress, showpacs, twocolumn]{revtex4-1}
\usepackage{graphicx}
\usepackage{dcolumn}

\makeatletter

\newcommand{\Rmnum}[1]{\expandafter\@slowromancap\romannumeral #1@}
\makeatother

\begin{document}

\title{Magnetic ordering with reduced cerium moments in hole-doped CeOs$_2$Al$_{10}$}

\author{D. D. Khalyavin}
\email{email: dmitry.khalyavin@stfc.ac.uk}
\affiliation{ISIS facility, STFC, Rutherford Appleton Laboratory, Chilton, Didcot, Oxfordshire, OX11-0QX,UK}
\author{D. T. Adroja}
\email{email: devashibhai.adroja@stfc.ac.uk}
\affiliation{ISIS facility, STFC, Rutherford Appleton Laboratory, Chilton, Didcot, Oxfordshire, OX11-0QX,UK}
\affiliation{Physics Department, University of Johannesburg, P.O. Box 524, Auckland Park 2006, South Africa}
\author{A. Bhattacharyya}
\affiliation{ISIS facility, STFC, Rutherford Appleton Laboratory, Chilton, Didcot, Oxfordshire, OX11-0QX,UK}
\affiliation{Physics Department, University of Johannesburg, P.O. Box 524, Auckland Park 2006, South Africa}
\author{A. D. Hillier}
\affiliation{ISIS facility, STFC, Rutherford Appleton Laboratory, Chilton, Didcot, Oxfordshire, OX11-0QX,UK}
\author{P. Manuel}
\affiliation{ISIS facility, STFC, Rutherford Appleton Laboratory, Chilton, Didcot, Oxfordshire, OX11-0QX,UK}
\author{A. M. Strydom}
\affiliation{Physics Department, University of Johannesburg, P.O. Box 524, Auckland Park 2006, South Africa}
\author{J. Kawabata}
\affiliation{Department of Quantum Matter, ADSM, and IAMR, Hiroshima University, Higashi-Hiroshima 739-8530, Japan}
\author{T. Takabatake}
\affiliation{Department of Quantum Matter, ADSM, and IAMR, Hiroshima University, Higashi-Hiroshima 739-8530, Japan}
\date{\today}

\begin{abstract}
The lightly hole-doped system CeOs$_{1.94}$Re$_{0.06}$Al$_{10}$ has been studied by muon spin relaxation and neutron diffraction measurements. A long-range antiferromagnetic ordering of the Ce-sublattice with substantially reduced value of the magnetic moment $0.18(1) \mu_B$ has been found below $T_N$ = 21 K. Similar to the undoped parent compound, the magnetic ground state of CeOs$_{1.94}$Re$_{0.06}$Al$_{10}$ preserves the anomalous direction of the ordered moments along the $c$-axis. The obtained result reveals the crucial difference between electron- and hole-doping effects on the magnetic ordering in CeOs$_2$Al$_{10}$. The former suppresses the anisotropic $c-f$ hybridization and promotes localized Ce moments. On the contrary, the latter increases the hybridization and shifts the system towards delocalized non-magnetic state.
\end{abstract}

\pacs{75.25.-j}

\maketitle

\indent Unusual magnetic order in the Kondo semiconductors CeOs$_2$Al$_{10}$ and CeRu$_2$Al$_{10}$ with an orthorhombic cage structure has attracted much attention in last few years.\cite{ref:1,ref:2,ref:3,ref:4,ref:5,ref:6,ref:7,ref:8,ref:9,ref:10,ref:11} The order takes place at unexpectedly high temperature $T_N\sim$ 29 K if one takes into account the large separation distance ($\sim 5.2 \AA$) between Ce ions in the structure and involvement of a very small ordered moment $\sim$ 0.3 - 0.4$\mu_B$ along the $c$-crystallographic direction\cite{ref:9,ref:12,ref:13} which is not the easy axis in the static susceptibility measurements above the transition temperature. In addition, a spin gap formation takes place at a temperature slightly higher $T_N$ in these materials as well as in the paramagnetic state of CeFe$_2$Al$_{10}$.\cite{ref:7,ref:11,ref:14,ref:15} The latter does not show any sign of magnetic order and the Ce ions in this compound are believed to be in the valence fluctuating regime. The mechanism providing the high ordering temperature and the gap formation has not been clarified yet. Based on optical conductivity measurements, Kimura et al.\cite{ref:16} suggested a charge density wave as the primary instability which then induces the unusual magnetic ordering at $T_N$. An indirect confirmation of this idea is the presence of the structural modulation with the $(0,-\frac{2}{3},\frac{2}{3})$ propagation vector revealed by electron diffraction in CeOs$_2$Al$_{10}$ at 15 K.\cite{ref:3} The critical temperature where this modulation comes in was not clarified and later X-ray and neutron diffraction measurements failed to find evidence of the symmetry lowering below and above $T_N$.\cite{ref:9,ref:11,ref:12}\\
\indent The reduced ordered moment found in the neutron diffraction experiments has been attributed by Strigari et al.\cite{ref:17,ref:18} to the crystal field effects studied by polarization-dependent soft x-ray absorption measurements. The proposed ground state wave function provides quantitative agreement with the experimental data and has been confirmed by inelastic neutron scattering.\cite{ref:19} The nature of the strongly anisotropic exchange parameters stabilizing the ordered state with the moments being along the $c$-axis is however unclear so far.\\
\indent Very recently, a great sensitivity of CeRu$_2$Al$_{10}$ and CeOs$_2$Al$_{10}$ to electron-doping has been revealed.\cite{ref:20,ref:21,ref:21a} In both compounds a small substitution with $4d$ and $5d$ transition metals carrying extra electrons results in a pronounced change of the magnetic structure. The doping induced ground state is controlled by the single-ion anisotropy and implies a considerably bigger ordered moment $\sim$ 0.92$\mu_B$ along the $a$-axis. The doping effect was attributed to a suppression of the anisotropic $c-f$ hybridization promoting localized states for the $4f$ electrons of Ce. In the light of this recent finding, the effect of a hole-doping is the natural step forward to investigate the anomalous magnetic order in this family of materials. In the present work, we studied the hole-doped system CeOs$_{1.94}$Re$_{0.06}$Al$_{10}$ by means of muon spin relaxation ($\mu^+$SR) and high resolution neutron diffraction. We found that this system preserves the anomalous direction of the ordered moments but their size is substantially reduced. This indicates that unlike electron-doped systems, the hole-doping increases the anisotropic $c-f$ hybridization shifting the system towards the non-magnetic delocalized state.\\
\indent The polycrystalline sample of CeOs$_{1.96}$Re$_{0.06}$Al$_{10}$ was prepared by arc melting of stoichiometric quantities of the starting elements with subsequent annealing at 850 C$^o$ for 6 days.  Zero-field muon spin-relaxation and powder neutron diffraction experiments were carried out at the ISIS pulsed neutron and muon facility of the Rutherford Appleton Laboratory, U.K., on the MuSR spectrometer and WISH diffractometer\cite{ref:22} on the second target station (TS-2). The $\mu ^+$SR experiments were conducted in longitudinal geometry. The powdered sample was mounted onto a 99.995 $+\%$ pure silver plate with Ge-varnish and was covered by a thin-Al foil for thermal equilibrium. 
\begin{figure}[t]
\includegraphics[scale=0.92]{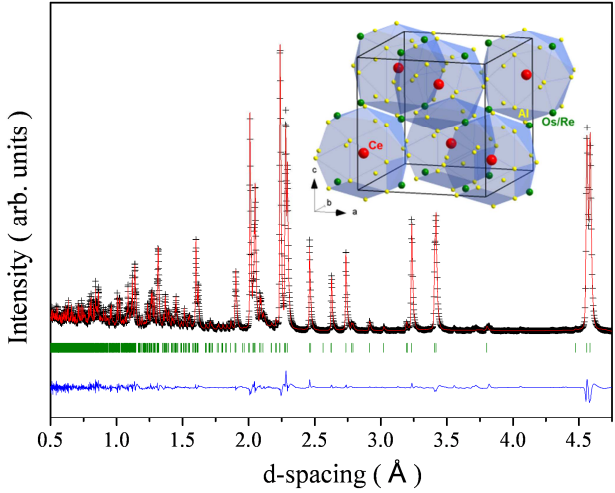}
\caption{(Color online) Rietveld refinement of the neutron powder diffraction pattern of  CeOs$_{1.94}$Re$_{0.06}$Al$_{10}$ collected at the backscattering detectors bank (average scattering angle is $154^o$) of the WISH diffractometer. The cross symbols and solid line (red) represent the experimental and calculated intensities, respectively, and the line below (blue) is the difference between them. Tick marks (green) indicate the positions of Bragg peaks in the $Cmcm$ space group. Inset shows schematic representation of the crystal structure of this composition as a stacking of the CeAl$_{16}$(Os/Re)$_4$ polyhedron cages.}
\label{fig:F1}
\end{figure}
\begin{table}[b]
\caption{Structural parameters of CeOs$_{1.94}$Re$_{0.06}$Al$_{10}$ refined from the neutron diffraction data collected at $T$ = 30 K in the orthorhombic $Cmcm$ ($a$=9.1217(1)$\AA$, $b$=10.2546(1)$\AA$, $c$=9.1704(1)$\AA$, $R_{Bragg}$=3.98 $\%$) space group. Occupancies for all the atoms in the refinement procedure were fixed to the nominal chemical content.}
\centering 
\begin{tabular*}{0.48\textwidth}{@{\extracolsep{\fill}} c c c c c c} 
\hline\hline\\ 
Atom & Site & $x$ & $y$ & $z$ & $B_{iso}$ \\ [1.5ex] 
\hline\\ 
Ce & 4$c$ & 0 & 0.1275(3) & 0.25 & 0.1(1)\\ 
Os/Re & 8$d$ & 0.25 & 0.25 & 0 & 0.38(3)\\
Al1 & 8$g$ & 0.2212(4) & 0.3640(4) & 0.25 & 0.1(1)\\
Al2 & 8$g$ & 0.3493(4) & 0.1329(4) & 0.25 & 0.3(1)\\
Al3 & 8$f$ & 0 & 0.1588(4) & 0.6019(4) & 0.6(1)\\
Al4 & 8$f$ & 0 & 0.3773(4) & 0.0493(4) & 0.5(1)\\
Al5 & 8$e$ & 0.2242(3) & 0 & 0 & 0.5(1)\\[1.5ex]
\hline
\hline  
\end{tabular*}
\label{table:T1} 
\end{table}
\begin{figure}[t]
\includegraphics[scale=0.75]{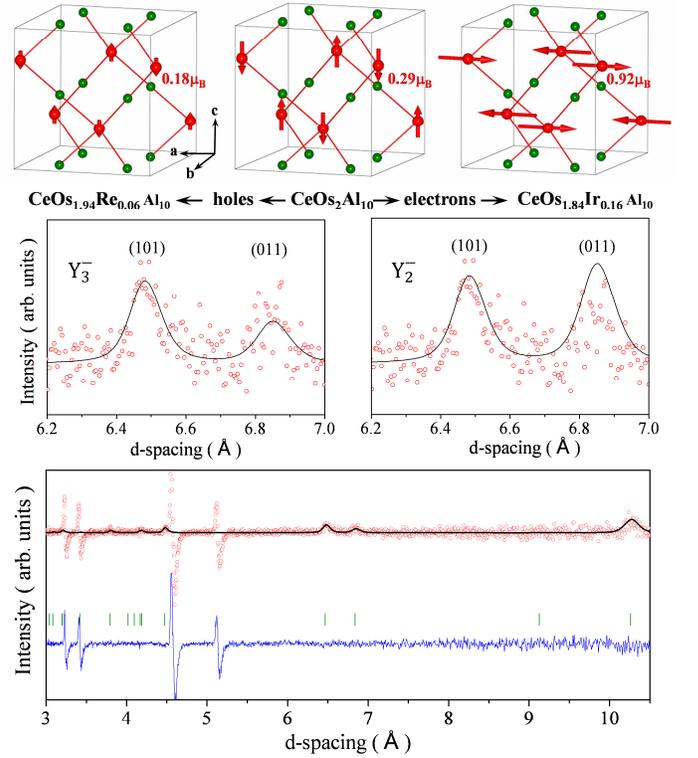}
\caption{(Color online) Magnetic structures of the hole-doped (found in this work), undoped\cite{ref:12} and electron-doped\cite{ref:21} samples. For clarity, only Ce and Os/Re atoms are shown (top). A part of the neutron diffraction patterns of CeOs$_{1.94}$Re$_{0.06}$Al$_{10}$ near the (101) and (011) magnetic reflections collected at 1.5 K (middle). Rietveld refinement of the magnetic intensity obtained as a difference between the diffraction patterns collected at 1.5 K and 30 K, in the model with the magnetic moments being along the $c$-axis. The circle symbols (red) and solid line represent the experimental and calculated intensities, respectively, and the line below (blue) is the difference between them. Tick marks (green) indicate the positions of Bragg peaks for the magnetic scattering with the (${\bf k}=1,0,0$) propagation vector.} 
\label{fig:F2}
\end{figure}
The sample and mount were then inserted into an Oxford Instrument (OI) cryostat and measured in the temperature range 1.2 K - 40 K. Muons implanting into any exposed part of the silver mount give rise to a flat time-independent background. For the neutron diffraction experiment, the sample (6g) was loaded into a cylindrical 6mm vanadium can and measured for ten hours at 1.5 K and 30 K using OI cryostat. The crystal and magnetic structure Rietveld refinements were performed using FullProf program\cite{ref:23} against the data measured in detector banks at average $2\theta $ values of $58^o$, $90^o$, $122^o$, and $154^o$, each covering $32^o$ of the scattering plane.\\
\indent The neutron diffraction pattern collected at $T$ = 30 K was successfully refined in the orthorhombic $Cmcm$ space group (Fig. \ref{fig:F1}) with the structural parameters summarized in Table \ref{table:T1}. The structure encapsulates the Ce ions into polyhedron cages formed by sixteen Al and four Os/Re atoms as shown in the inset of Fig. \ref{fig:F1}. This structural model works also well for the low temperature pattern collected at $T$ = 1.5 K. At this temperature, however, several extremely weak magnetic reflections have been observed in the $d$-spacing range above $3 \AA$. Similar to the undoped CeOs$_2$Al$_{10}$, these reflections can be indexed with the propagation vector ${\bf k}=(1,0,0)$ indicating antiferromagnetic ordering with the doubled primitive unit cell along the $a$- and $b$-directions (${\bf k_{prim}}=\frac{1}{2},\frac{1}{2},0$).\cite{ref:9,ref:12} The refinement procedure with such small magnetic signal is not straightforward and it is difficult to determine precisely the moments direction. Assuming irreducible nature of the magnetic order parameter, the appropriate symmetry restrictions were included in the analysis, which allowed us to reduce significantly the number of possible magnetic structures. Between all the tested models, two magnetic configurations with the symmetries of the $Y^-_3$ and $Y^-_2$ irreducible representations of the $Cmcm$ space group \cite{ref:24,ref:25} yield almost equally good fitting (Fig. \ref{fig:F2} (middle)). These models imply ordering of the Ce-sublattice along the $c$- and $a$-axes, respectively, as observed in the undoped CeOs$_2$Al$_{10}$ and the electron-doped CeOs$_{1.86}$Ir$_{0.16}$Al$_{10}$.\cite{ref:12,ref:21} Both configurations introduce a strongly reduced ordered moment $\sim$  0.18(1)$\mu_B$ per Ce ion. A close inspection of the observed and calculated magnetic intensities revealed overestimation of the (0,1,1) reflection in the model with the moments being along the $a$-direction, whereas the model with the moments along the $c$-axis reproduces all reflections accurately (Fig. \ref{fig:F2} (bottom)). 
\begin{figure}[t]
\includegraphics[scale=1.03]{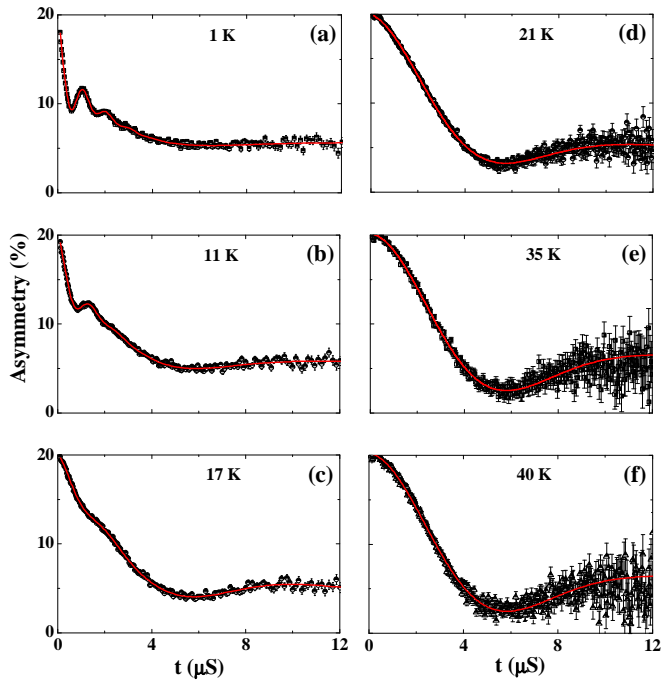}
\caption{(Color online) The muon depolarization spectra of CeOs$_{1.94}$Re$_{0.06}$Al$_{10}$ measured at different temperatures. The line is a least-squares fit of the data using Eqs.\ref{eq:F1} (right) and \ref{eq:F2} (left)}.
\label{fig:F3}
\end{figure}
\begin{figure}[t]
\includegraphics[scale=0.50]{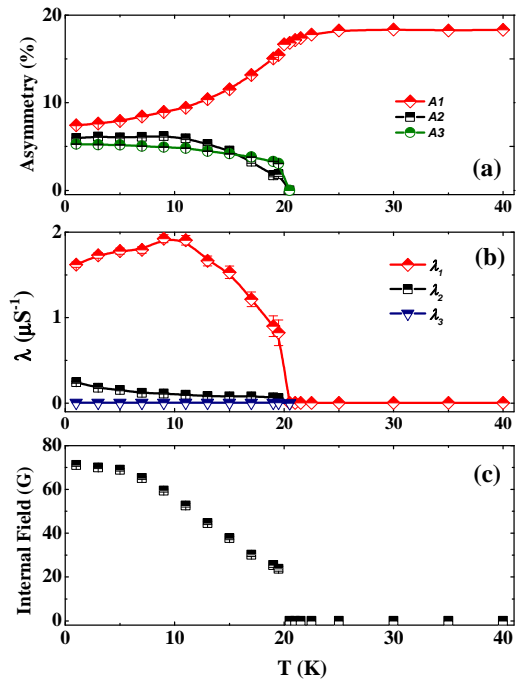}
\caption{(Color online) Temperature  dependencies of the asymmetry (a), relaxation rate (b) and the internal field on the $4c$ muon stop site (c) obtained from the fitting of the $\mu^+$SR spectra in Fig. \ref{fig:F3} using Eq.\ref{eq:F2}} 
\label{fig:F4}
\end{figure}
Thus, the latter configuration is the most probable indicating that the hole-doping preserves the anomalous direction of the ordered moments but drastically decreases their size (Fig. \ref{fig:F2} (top)). A temperature dependence of the magnetic order parameter requires an unreasonable amount of beam-time and was not measured in the present neutron diffraction study. The information about the transition temperature has been obtained from the zero-field $\mu^+$SR measurements which confirmed the long-range ordered nature of the magnetic ground state in CeOs$_{1.94}$Re$_{0.06}$Al$_{10}$.\\
\indent Figure \ref{fig:F3} shows the zero-field $\mu^+$SR spectra collected in the temperature range of 1.2 K $<T<$ 40 K. Between 21 and 40 K a strong damping at shorter time and the recovery at longer times have been observed, which is the typical muon response to a static distribution of the nuclear dipole moments, described by the Kubo-Toyabe (KT) formalism.\cite{ref:26} This indicates that above 21 K, CeOs$_{1.94}$Re$_{0.06}$Al$_{10}$ is a paramagnet and the $\mu^+$SR spectra in this temperature range were analyzed using the equation:
\begin{eqnarray}
G_zKT(t)=A_1exp (-\lambda_1 t) \nonumber \\ \left (  \frac{1}{3}+\frac{2}{3}(1-(\sigma_{KT}t)^2) exp\left ( -\frac{(\sigma_{KT}t)^2}{2} \right )\right )+BG 
\label{eq:F1}
\end{eqnarray}
where $A_1$ is the initial asymmetry, $\sigma_{KT}$ is nuclear depolarization rate, $\sigma_{KT}/\gamma_{\mu }=\Delta$ is the local Gaussian field distribution width, $\gamma_{\mu }$ is the gyromagnetic ratio of the muon, $\lambda_1$ is the electronic relaxation rate and BG is a constant background arising from muon stopping on the Ag-sample holder. The BG was estimated from 40 K data and was kept fixed for fitting all the other spectra. The value of $\sigma_{KT}$ was obtained to be $\sim$ 0.30 $\mu s^{-1}$ from fitting the spectra above 21 K and this value was found to be temperature independent down to 1.2 K. Using a similar value for $\sigma_{KT}$, Kambe et al\cite{ref:10} suggested $4a(0,0,0)$ as the muon stopping site in CeRu$_2$Al$_{10}$, while for CeOs$_2$Al$_{10}$ the stopping site was assigned to $4c(0.5,0,0.25)$.\cite{ref:11} Recently Guo et al\cite{ref:27} have investigated Ce(Ru$_{1-x}$Rh$_x$)$_2$Al$_{10}$ ($x=0-0.08$) series using muon spin relaxation and their dipolar fields calculation supports the $4c$ position, while for the $4a$ one, these authors found almost zero internal field.\\
\indent Below 21 K, the $\mu^+$SR spectra of CeOs$_{1.94}$Re$_{0.06}$Al$_{10}$  reveal coherent frequency oscillations indicating a long range magnetic ordering (Fig. \ref{fig:F3} (a-c)). The spectra can be described by one oscillatory term with exponential decay plus KT term, as given by the equation:
\begin{eqnarray}
G_z(t)=G_z KT(t)+A_2 cos(\omega t+\varphi)\nonumber \\ 
exp(-\lambda _2t)+A_3 exp(-\lambda _3t)
\label{eq:F2}
\end{eqnarray}
where $\omega=\gamma_{\mu }H_{int}$ is the muon precession frequency ($H_{int}$ is the internal field at the muon site), $\lambda _i$ ($i$=2 and 3) is the muon depolarization rate (arising from the distribution of the internal field) and $\varphi $ is the phase. The second term represents the transverse components of the internal fields seen by the muons along which they precess, while the third term represents the longitudinal component.  The first term in Eq.(\ref{eq:F2}) is the KT term, which accounts for the dip seen in the $\mu^+$SR spectra near 5 $\mu s$ even in the magnetically ordered state, indicating that the internal field seen by the muons on this site is smaller than that of the nuclear field. This suggests that the corresponding muon site can be assigned to the $4a$ crystallographic position.\cite{ref:27}\\
\indent Fig. \ref{fig:F4} (a-c) shows the temperature dependence of the fitting parameters obtained using the equation (\ref{eq:F2}). One can see that asymmetry $A_1$ associated with KT term is temperature independent between 40 K and 21 K and exhibits nearly 2/3 drop below 21 K (Fig. \ref{fig:F4} (a)) indicating that the long-range magnetic ordering occurs in the full volume of the sample and hence has a bulk nature. The asymmetries $A_2$ and $A_3$ appear below 20 K and exhibit saturation at low temperatures (Fig. \ref{fig:F4} (a)). On the other hand the relaxation rate, $\lambda_1$ associated with the KT term demonstrates a sharp rise below 20 K and a maximum near 10 K, while $\lambda_2$ and $\lambda _3$ show weak temperature dependence (Fig. \ref{fig:F4} (b)). The muon precession frequency as a function of temperature indicates that the internal field, for the muon stopping on the $4c$ site, appears just below 20 K (Fig. \ref{fig:F4} (c)). The muons stopping on the $4a$ site experience almost zero-internal field, which is consistent with the fact that the $Y^-_3$ irreducible representation transforming the magnetic order parameter has zero subduction frequency for the associated reducible magnetic representation on this site. This means that the vanishingly small internal field on the $4a$ site is the direct consequence of the magnetic symmetry which fully compensates the molecular fields on this site. The subduction frequency of $Y^-_3$ for the magnetic representation on the $4c$ site is non-zero and therefore a finite external field is expected for this site by symmetry, in agreement with the experimental data. The small value of the internal fields observed through $\mu^+$SR study is consistent with the reduced ordered moment found using neutron diffraction. \\
\indent Thus the muon spin relaxation and neutron diffraction both indicate the magnetic nature of the ground state in the hole-doped system CeOs$_{1.94}$Re$_{0.06}$Al$_{10}$. The obtained result, however, reveals the fundamental difference between the electron and hole-doping effects on the magnetic properties of CeOs$_2$Al$_{10}$. The electron-doping suppresses the anisotropic $c-f$ hybridization and promotes localized states controlled by the single ion anisotropy with the big ordered moment 0.92$\mu_B$. On the contrary, hole-doping increases the hybridization reducing the moment and shifting the system towards the non-magnetic state. This great sensitivity of the magnetic ground state of CeOs$_2$Al$_{10}$ to the carrier type and concentration allows a continuous tuning of the system from the magnetic localized to non-magnetic delocalized state. A similar control of the magnetic ground state with the carriers-doping has been observed before in the non-magnetic Kondo semimetal CeNiSn, where a substitution of Cu for Ni induced a long-range antiferromagnetic order of the Ce moments.\cite{ref:28} \\
\indent In conclusion, the lightly hole-doped system CeOs$_{1.96}$Re$_{0.06}$Al$_{10}$ exhibits antiferromagnetic ordering of the Ce-moments below $T_N\sim$ 21 K. The ordered moments are substantially reduced but preserve the anomalous direction along the $c$-axis similar to the undoped compound. Enhancement of the $c-f$ hybridization with the hole concentration is assumed to be the reason of the moment reduction. This indicates the fundamental difference between the hole- and electron-doping effects on the magnetic ordering in CeOs$_2$Al$_{10}$ and allows tuning the magnetic ground state from the ordered localized to the non-magnetic delocalized.\\ 
\indent Acknowledgment: We would like to thank Peter Baker for interesting discussions. DTA/ADH would like to acknowledge financial assistance from CMPC-STFC Grant No. CMPC-09108. The work at Hiroshima University was supported by a Grant-in-Aid for Scientific Research on Innovative Area "Heavy Electrons" (20102004) of MEXT, Japan. AB would like to thank JU and STFC for a PDF funding. AMS thanks the SA-NRF (78832) for financial assistance. 

\end{document}